\documentclass[11pt]{article}

\usepackage[preprint]{acl}
\usepackage{times}
\usepackage{latexsym}
\usepackage[T1]{fontenc}
\usepackage[utf8]{inputenc}
\usepackage{microtype}
\usepackage{inconsolata}
\usepackage{graphicx}
\usepackage{booktabs}
\usepackage{array}
\usepackage{multirow}
\usepackage{amsmath}
\usepackage{amssymb}
\usepackage{xcolor}
\usepackage{pifont}
\usepackage{placeins}

\graphicspath{{figures/}}

\newcommand{\method}{\textsc{Evolution Under Fire}}

\definecolor{eufcoral}{HTML}{E25867}
\definecolor{eufblue}{HTML}{4D87A8}
\definecolor{eufblend}{HTML}{987088}
\newcommand{\algstage}[2][eufcoral]{\textcolor{#1}{\textbf{#2}}}

\newcounter{eufalgorithm}
\newenvironment{eufalgorithm}[1]{%
  \refstepcounter{eufalgorithm}%
  \par\medskip\noindent
  \begin{minipage}{\columnwidth}
  \hrule height 0.65pt \vspace{2pt}
  \textbf{Algorithm \theeufalgorithm\quad #1}\par\vspace{2pt}
  \hrule height 0.30pt \vspace{3pt}
  \begin{tabular}{@{}r@{:\hspace{0.55em}}>{\raggedright\arraybackslash}p{0.90\linewidth}@{}}
}{%
  \end{tabular}
  \vspace{2pt}\hrule height 0.65pt
  \end{minipage}
  \par\medskip
}

\title{Evolution Under Fire: Evolving Agent Safety Harnesses with Adaptive Injection Stressors}

\author{
\textbf{Yujun Wang\textsuperscript{1,2}}\thanks{\raggedright Corresponding authors: \texttt{yujun\_wang\_cn@hotmail.com}, \texttt{cognitive.yunpu@gmail.com}.} \quad
\textbf{Yilun Liu\textsuperscript{1,2}} \quad
\textbf{Jinhe Bi\textsuperscript{1,2}} \quad
\textbf{Krishna Kanth Nakka} \\
\textbf{Xuebing Zhou} \quad
\textbf{Hinrich Sch{\"u}tze\textsuperscript{1,2}} \quad
\textbf{Yunpu Ma\textsuperscript{1,2}}\footnotemark[1] \\[0.45em]
\normalfont
\textsuperscript{1}LMU Munich; \quad
\textsuperscript{2}Munich Center for Machine Learning (MCML)
}

\begin{document}
\maketitle

\begin{abstract}
Tool-using language-model agents must act on information from untrusted emails, webpages, and documents.
This creates a central security challenge: malicious instructions and legitimate task directives can arrive through the same interaction channel.
Existing defenses are largely hand-designed and static, making further improvements dependent on repeated manual redesign.
We introduce Evolution Under Fire, a staged framework that couples safety-harness evolution with adaptive stressor evolution.
Utility-constrained, Pareto-scheduled tree search evolves complete executable safety harnesses from execution evidence while keeping the task model, benchmark, tools, and evaluator fixed.
Adaptive stressor evolution then discovers utility-preserving attacks that guide a subsequent stage of harness evolution.
On AgentDyn with GPT-4o-mini, an evolved harness reduces attack success from 50.4\% to 3.0\%, improves clean and attacked utility over the initial harness, and offers a stronger security--utility trade-off than reproduced defenses.
On Agent-SafetyBench with Qwen2.5-7B-Instruct, harness evolution reduces attack success from 82.6\% to 62.0\% while retaining helpfulness and runtime validity.
Evolved harnesses also transfer without further evolution to Qwen3.5-35B-A3B, reducing attack success from 32.1\% to 0.9\% while improving attacked-task utility.
\end{abstract}

\section{Introduction}
\label{sec:introduction}

\begin{figure}[t!]
  \centering
  \includegraphics[width=\columnwidth]{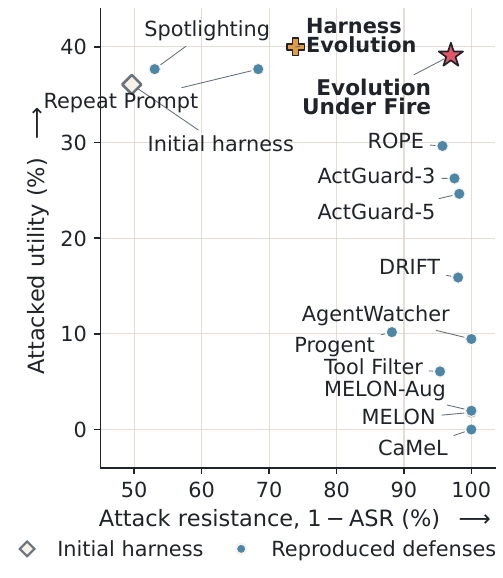}
  \caption{\textbf{Security--utility trade-off on AgentDyn.}
  \method{} combines near-maximal attack resistance with strong attacked-task utility, occupying the favorable frontier across both objectives.}
  \label{fig:security-utility}
\end{figure}

\begin{figure*}[t!]
  \centering
  \includegraphics[width=0.90\textwidth]{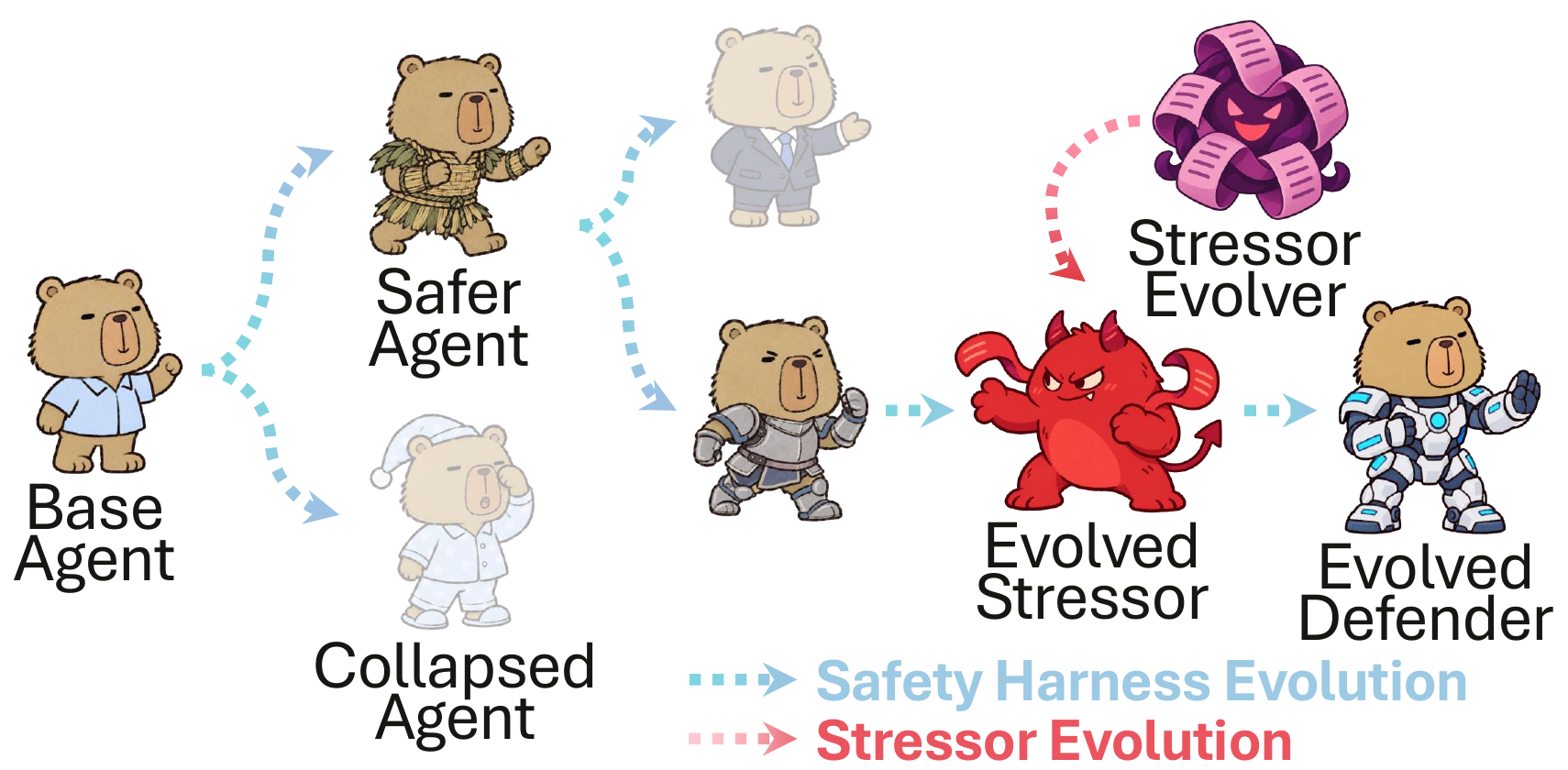}
  \caption{\textbf{Illustration of \method.}
  Efficient harness evolution improves agent defenses, while adaptive stressor evolution exposes them to increasingly challenging attacks.
  Their feedback drives the search toward agents that remain capable and robust under adversarial pressure.}
  \label{fig:teaser}
\end{figure*}

Language-model agents increasingly complete user requests by reading emails, webpages, repositories, and other external content before invoking consequential tools.
The same interface that makes these agents useful also creates an indirect prompt-injection channel: an attacker can place instructions in data that the agent later treats as authority \citep{greshake2023not,agentdojo2024,injecagent2024}.
The resulting failure is not confined to an unsafe response.
It can change a tool call, disclose private data, or commit an external side effect.

Defending this setting exposes a difficult security--utility trade-off.
Prompt reminders and content filters can suppress attacks but also discard instructions that are legitimately required by open-ended tasks.
Conversely, permissive agents preserve utility while remaining vulnerable.
Recent system-level defenses introduce planning constraints, information-flow rules, privilege policies, causal monitors, or action auditors \citep{camel2026,progent2025,drift2025,melon2025,agentwatcher2026,actguard2026,rope2026}.
Each design encodes a useful hypothesis, but each is a fixed point in a large program space: the surrounding harness can choose how to construct context, expose tools, maintain state, invoke auxiliary models, validate actions, and recover from uncertainty.

At the same time, a fixed attack suite is an incomplete optimization signal.
AgentDojo provides executable tasks and security oracles \citep{agentdojo2024}, while AgentDyn adds open-ended tasks whose legitimate completion can depend on external instructions \citep{agentdyn2026}.
Yet repeatedly optimizing against the same attack family risks producing benchmark-specific defenses.
Adaptive attack work shows that static evaluations can overestimate defense efficacy when attacks are adapted to the target defense \citep{autodojo2026}.
This suggests that neither side should remain a single hand-designed artifact: the safety harness should improve from its failures, and the attacks used to probe it should improve as well.

This raises the central question of this paper:
\begin{quote}
\emph{How can executable safety harnesses evolve from observed failures, and how can adaptive, utility-preserving attacks further strengthen their defenses?}
\end{quote}

\begin{figure*}[t!]
  \centering
  \includegraphics[width=\textwidth]{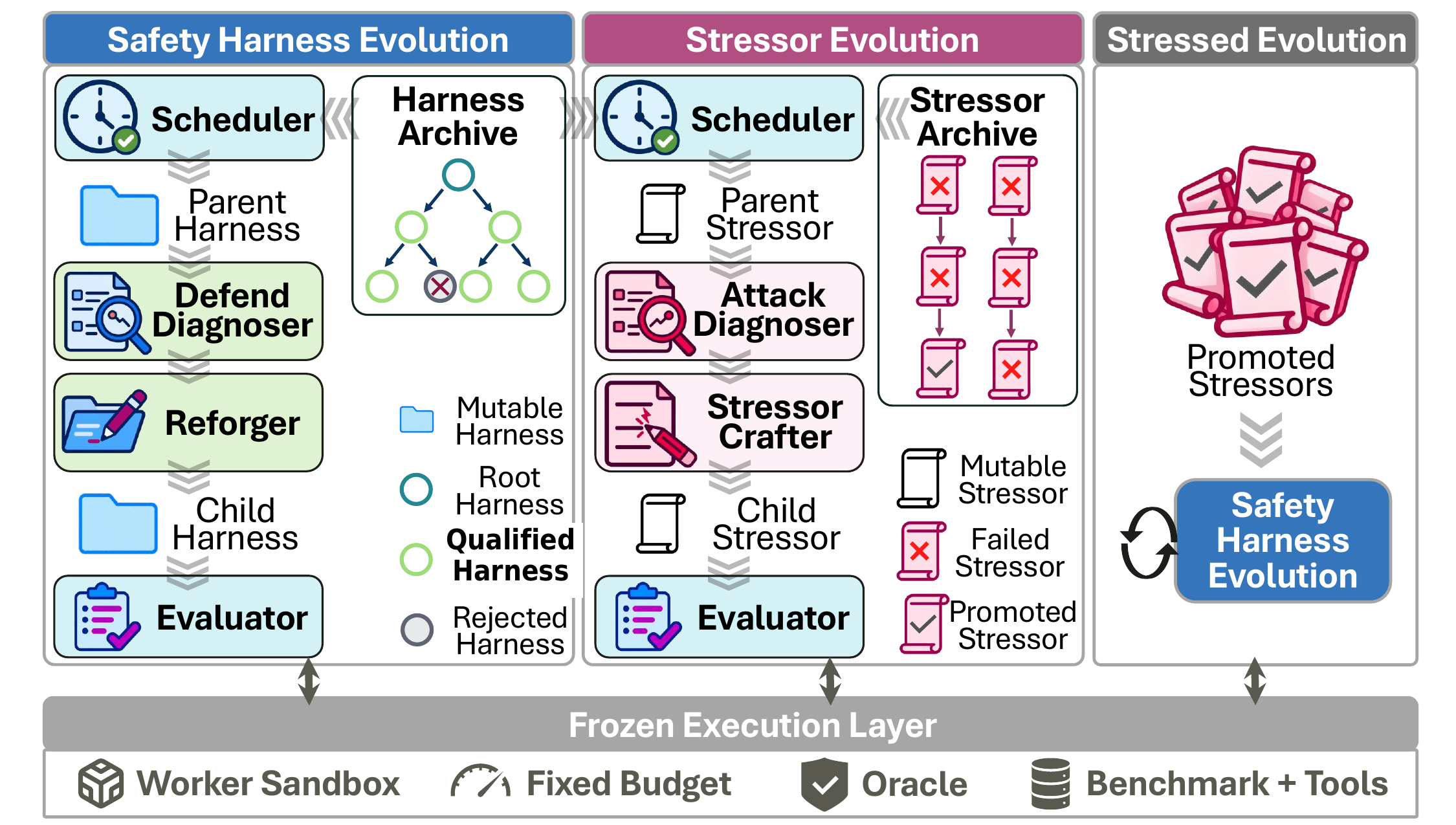}
  \caption{\textbf{Overview of \method.}
  Harness evolution first searches for utility-preserving defenses.
  Stressor evolution then targets residual weaknesses in robust harnesses, and verified stressors provide feedback for subsequent harness evolution.
  The task model, benchmark, tools, and evaluator remain fixed throughout.}
  \label{fig:pipeline}
\end{figure*}

In response to this question, we propose \method, a framework that places safety-harness evolution under adaptively generated injection pressure while keeping the task model and benchmark fixed.
The defense-side process is \emph{utility-constrained, Pareto-scheduled harness tree evolution}.
Each expansion selects one utility-qualified parent and produces one child harness, while a scheduler can revisit any previously evaluated branch rather than committing to a linear chain.
Candidates that violate utility constraints cannot become parents, but remain visible as negative evidence.
The attack-side process is \emph{utility-preserving, lineage-based stressor evolution}.
It opens or continues stressor lineages against cases a target harness originally solved securely, and retains a generated stressor as feedback only when the attack succeeds without destroying the benign task.

Together, the two searches form a staged adversarial bootstrapping process: evolved harnesses guide the discovery of stronger stressors, which in turn strengthen subsequent harness evolution.

Our contributions are:
\begin{itemize}
    \item \textbf{Executable safety-harness evolution.}
    We formulate agent defense as search over complete executable harnesses within a fixed evaluation environment.
    A read-only Diagnoser and code-generating Expander ground each edit in execution traces and archive-wide search history.
    \item \textbf{Utility-constrained Pareto tree search.}
    Harness evolution admits only candidates that preserve utility relative to the initial harness, then samples parents from the Pareto frontier using uncertainty-aware metric estimates.
    It maintains diverse security--utility trade-offs, allows strong older branches to receive further expansions, and retains utility-rejected candidates as counterexamples without allowing them to propagate.
    \item \textbf{Feedback from adaptive stressors.}
    Our stressor evolution generates adaptive stressors whose outcomes are verified by benchmark oracles.
    These adaptive stressors expose residual weaknesses in strong harnesses and provide targeted feedback for subsequent defense evolution.
    Stressor-guided evolution improves the security--utility frontier over a budget-matched harness-only control, and the resulting defenses transfer across closed and open-weight task models.
\end{itemize}

\section{Related Work}
\label{sec:related}

\paragraph{Prompt-injection defenses for agents.}
Agent defenses mark or filter untrusted content, constrain privileges and information flow, or monitor proposed actions.
Representative systems include CaMeL, Progent, DRIFT, MELON, AgentWatcher, ActGuard, and ROPE \citep{camel2026,progent2025,drift2025,melon2025,agentwatcher2026,actguard2026,rope2026}.
Whereas these methods instantiate particular defense mechanisms, \method{} searches over executable harnesses under utility constraints.
Appendix~\ref{app:baseline-details} describes the comparison methods and reproduction settings.

\paragraph{Automated agent and harness design.}
Meta-optimization has evolved agent programs and workflows through archive- or lineage-based search \citep{adas2025,dgm2026,hgm2026}.
Meta-Harness and Self-Harness ground revisions in source, traces, and observed failures, while SHE evolves structured safety-harness components \citep{metaharness2026,selfharness2026,she2026}.
\method{} instead combines utility-constrained tree search over complete executable harnesses with feedback from adaptive stressors.

\paragraph{Automated and adaptive red teaming.}
Automated red teaming uses iterative refinement, branching search, and quality-diversity optimization to discover adversarial inputs \citep{pair2025,tap2024,rainbow2024}.
AutoDojo adapts injections to target tool-using agents \citep{autodojo2026}; \method{} additionally feeds verified, utility-preserving stressors back into harness evolution.

\section{Problem Setting}
\label{sec:problem}

Let $M$ be a frozen task model and $h$ an executable harness that controls how $M$ receives context, invokes tools, validates actions, maintains state, and recovers from failures.
For benchmark case $x$, the trusted environment executes the composed agent $h[M]$ and returns utility and security outcomes.
The benchmark, task data, tools, environment state, oracles, and evaluator are fixed throughout evolution.

\paragraph{Multi-objective evaluation.}
Let $\mathbf{U}(h)$ denote the benchmark-defined utility outcomes to maximize and $\mathbf{R}(h)$ the security-risk outcomes to minimize.
When adaptive stressors are introduced, their task-preservation and attack outcomes augment these objective vectors.
All coordinates remain separate rather than being collapsed into a scalar reward.

\paragraph{Utility-constrained objective.}
We seek harnesses that minimize security risk subject to benchmark-defined utility requirements.
This constraint rules out solutions that obtain apparent safety by disabling useful behavior.

\paragraph{Trusted boundary.}
Only the candidate harness is mutable.
Model requests and tool actions pass through evaluator-controlled interfaces under fixed resource limits.
The Diagnoser can inspect evidence and the Expander can modify the selected candidate, but neither can alter the benchmark, oracles, evaluator, or recorded outcomes.

\section{Evolution Under Fire}
\label{sec:method}

\subsection{Overview}

\method{} proceeds in three stages.
It first evolves utility-qualified harnesses against a fixed benchmark, then evolves adaptive stressors that expose residual weaknesses in robust harnesses without changing the benchmark semantics.
The verified stressors subsequently provide additional evidence and objectives for further harness evolution.
Figure~\ref{fig:pipeline} summarizes this feedback-coupled process.

\begin{figure}[t]
\begin{eufalgorithm}{Harness Under Fire}\label{alg:method-overview}
1 & \textbf{Input:} benchmark $\mathcal{B}$, initial harness $h_0$, and search budgets $B_1,B_S,B_2$. \\
2 & \algstage[eufblue]{Harness evolution:} Evaluate $h_0$ and initialize the harness archive $\mathcal{H}$. \\
3 & \textbf{repeat} $B_1$ times: \\
4 & \quad Select a utility-qualified parent by posterior-Pareto sampling. \\
5 & \quad Diagnose its evidence, expand one child, evaluate it, and update $\mathcal{H}$. \\
6 & \algstage{Stressor evolution:} Form targets $\mathcal{T}$ from capable, secure harness--case pairs. \\
7 & \textbf{repeat} $B_S$ times: \\
8 & \quad Select a target or stressor lineage; diagnose, generate, and evaluate one descendant. \\
9 & Promote verified stressors that preserve task utility and achieve the adversarial objective. \\
10 & Deduplicate promoted stressors to form $\mathcal{S}_{\mathrm{adapt}}$. \\
11 & \algstage[eufblend]{Stressor feedback:} Evaluate utility-qualified harnesses on $\mathcal{S}_{\mathrm{adapt}}$. \\
12 & \textbf{repeat} $B_2$ times: evolve $\mathcal{H}$ using benchmark and adaptive-stressor objectives. \\
13 & \textbf{Return:} utility-qualified Pareto harnesses. \\
\end{eufalgorithm}
\end{figure}

\begin{table*}[!t]
\centering
\begin{tabular}{lcccccccc}
\toprule
\multirow{2}{*}{Method}
& \multicolumn{4}{c}{GPT-4o-mini}
& \multicolumn{4}{c}{Qwen3.5-35B-A3B} \\
\cmidrule(lr){2-5}\cmidrule(lr){6-9}
& CU $\uparrow$ & VU $\uparrow$ & AU $\uparrow$ & ASR $\downarrow$
& CU $\uparrow$ & VU $\uparrow$ & AU $\uparrow$ & ASR $\downarrow$ \\
\midrule
Initial harness
& 27/60 & 28/28 & 202/560 & 282/560
& 41/60 & 27/28 & 341/560 & 180/560 \\
Repeat User Prompt
& 25/60 & 27/28 & 211/560 & 177/560
& 42/60 & 26/28 & 373/560 & 97/560 \\
Spotlighting
& 25/60 & 25/28 & 211/560 & 263/560
& 41/60 & 26/28 & 347/560 & 159/560 \\
Tool Filter
& 3/60 & 14/28 & 34/560 & 26/560
& 1/60 & 7/28 & 17/560 & 0/560 \\
DRIFT
& 9/60 & 19/28 & 89/560 & 11/560
& 0/60 & 7/28 & 4/560 & 0/560 \\
AgentWatcher
& 23/60 & 28/28 & 53/560 & 0/560
& 27/60 & 23/28 & 68/560 & 1/560 \\
ActGuard ($k{=}3$)
& 26/60 & 22/28 & 147/560 & 14/560
& 6/60 & 12/28 & 47/560 & 0/560 \\
ActGuard ($k{=}5$)
& 24/60 & 22/28 & 138/560 & 10/560
& 8/60 & 13/28 & 47/560 & 0/560 \\
MELON
& 3/60 & 17/28 & 10/560 & 0/560
& 3/60 & 19/28 & 49/560 & 6/560 \\
MELON-Aug
& 2/60 & 17/28 & 11/560 & 0/560
& 8/60 & 18/28 & 48/560 & 2/560 \\
Progent
& 3/60 & 18/28 & 57/560 & 66/560
& 4/60 & 18/28 & 86/560 & 510/560 \\
ROPE
& 18/60 & 24/28 & 166/560 & 24/560
& 27/60 & 20/28 & 258/560 & 25/560 \\
CaMeL
& 0/60 & 13/28 & 0/560 & 0/560
& 0/60 & 13/28 & 0/560 & 0/560 \\
\midrule
Ours
& 28/60 & 25/28 & 219/560 & 17/560
& 39/60 & 25/28 & 355/560 & 5/560 \\
\bottomrule
\end{tabular}
\caption{Complete reproduced AgentDyn comparison under two task models. CU is clean utility, VU validator utility, AU attacked utility, and ASR is attack success rate. ROPE uses the GPT-5 router. The reported \method{} rows are selected balanced harnesses; the Qwen3.5-35B-A3B result is transferred without further evolution. Values are exact successful-case counts and denominators.}
\label{tab:main-results}
\end{table*}

\subsection{Utility-Constrained Harness Tree Evolution}

This stage searches for stronger executable harnesses without accepting apparent safety gains obtained by sacrificing task completion.
Each expansion selects a previously evaluated, utility-qualified harness, diagnoses its behavior in the context of earlier attempts, generates one modified harness, and evaluates the result.
Repeating this loop produces a tree-structured search over alternative security--utility trade-offs rather than a single sequence of accepted edits.

\paragraph{Branching harness search.}
The search begins from an initial harness and organizes all evaluated candidates as a rooted tree $\mathcal{H}$.
Each child $h_c$ inherits its implementation from one selected parent $h_p$ and records its evaluation outcomes, qualification status, and supporting evidence.
The parent relation therefore captures implementation ancestry, while the scheduler may later return to any utility-qualified branch.

\paragraph{Evidence-grounded proposals.}
For the selected parent, a read-only Diagnoser examines its source, summarized metrics, and execution traces together with relevant evidence from the rest of the search tree.
It reports observations, trade-offs, supporting evidence, and uncertainties.
A separate Expander uses this diagnosis and the available evidence to produce the child harness, while remaining free to choose its architecture and implementation.
Candidates that fail utility or execution requirements remain visible as counterexamples, but cannot become parents.
Thus a child inherits code from one parent while learning from the broader search history.

\paragraph{Utility qualification.}
Evaluation proceeds from inexpensive checks to increasingly comprehensive tests.
Candidates that fail execution or utility requirements are rejected early, while survivors receive complete utility and security evaluation, including evaluation against adaptive stressors when available.
Only complete and valid outcomes can establish utility qualification for future expansion.

\paragraph{Pareto parent selection.}
Among utility-qualified harnesses, the scheduler accounts for both objective trade-offs and uncertainty from finite evaluation counts.
For each utility-qualified harness $h$ and Bernoulli objective $j$, the scheduler samples
\begin{equation}
  \tilde{p}_{h,j} \sim \mathrm{Beta}(s_{h,j}+1, n_{h,j}-s_{h,j}+1),
\end{equation}
where $s_{h,j}$ successes are observed in $n_{h,j}$ cases.
Attack objectives are sign-reversed for comparison.
The scheduler forms the first non-dominated front under the sampled vector and chooses a parent from that front.
Posterior sampling reflects uncertainty in the observed success rates, while Pareto selection retains candidates representing different security--utility trade-offs.

\subsection{Adaptive Stressor Evolution}

Harness evolution against a fixed collection of attacks can eventually exhaust the available failure signal.
This stage therefore searches for new benchmark-compatible stressors against harnesses that already behave both usefully and securely.
It modifies only the benchmark-defined adversarial input while keeping the user task, attacker objective, tools, environment, and evaluation criteria fixed.
The resulting stressors expose residual weaknesses without changing what counts as successful task completion or attack success.

\paragraph{Selecting informative targets.}
The search targets harness--case pairs for which the harness completes the user task and remains secure under the benchmark's existing adversarial condition.
Such pairs provide meaningful starting points: the harness is currently capable and secure, so a successful new stressor reveals a previously unseen weakness.
At each expansion, the scheduler either opens a new target pair to broaden coverage or continues an existing stressor lineage to refine earlier attempts.

\paragraph{Evolving stressor lineages.}
A lineage begins from the benchmark-provided adversarial instance for a target case and may continue through generated descendants.
For a selected harness $h$, case $x$, and stressor parent $z_p$, a read-only Stressor Diagnoser examines the target execution and relevant successful and failed stressors.
A Stressor Generator then produces one child $z_c$, which is evaluated through the attack surface defined for $x$.
Only the stressor changes; the surrounding benchmark case and its evaluation semantics remain fixed.

\paragraph{Utility-preserving promotion.}
Let $U(h,x,z)$ indicate successful completion of the user task and $A(h,x,z)$ indicate success of the adversarial objective.
A generated stressor is promoted only when
\begin{equation}
  \mathrm{Promote}(z_c)=U(h,x,z_c)\,A(h,x,z_c)=1,
\end{equation}
with complete, error-free execution.
Thus the adversarial objective must succeed while the original user task remains successful.
All evaluated stressors remain available as search evidence, but unsuccessful or utility-damaging stressors are not promoted.
Promoted stressors are deduplicated before forming the set.

\subsection{From Stressors Back to Harness Evolution}

Promoted stressors form the adaptive stressor set $\mathcal{S}_{\mathrm{adapt}}$ for the next harness-evolution stage.
Task preservation on $\mathcal{S}_{\mathrm{adapt}}$ becomes a qualification requirement, attack success on $\mathcal{S}_{\mathrm{adapt}}$ becomes an additional Pareto objective, and representative outcomes enter the search evidence.
In this way, robust harnesses guide stressor discovery, and the resulting stressors redirect subsequent harness search toward residual weaknesses.

\section{Experimental Setup}
\label{sec:experiments}

\paragraph{Benchmarks.}
We evaluate on AgentDyn \citep{agentdyn2026}, which jointly measures task
completion and resistance to indirect prompt injection, and Agent-SafetyBench
\citep{agentsafetybench2024}, which covers broader behavioral risks across
interactive agent environments.  Appendix~\ref{app:benchmark-details} gives
the versions, task inventories, and evaluation semantics.

\paragraph{Models.}
AgentDyn uses GPT-4o-mini and Agent-SafetyBench uses Qwen2.5-7B-Instruct as
the task model.  All evolution roles use GPT-5 through OpenRouter and remain
separate from the evaluated agent.  We test transfer by applying frozen
evolved harnesses to Qwen3.5-35B-A3B without further evolution.

\paragraph{Metrics.}
For AgentDyn, we report clean utility (CU, $\uparrow$), validator utility
(VU, $\uparrow$), attacked utility (AU, $\uparrow$), and attack success rate
(ASR, $\downarrow$); the adaptive stressor set is evaluated with utility and
ASR.  For Agent-SafetyBench, ASR is the complement of its Safety Score.  Since
that benchmark has no task-utility metric, we additionally measure
Helpfulness and runtime validity.  We report the coordinates
separately rather than optimizing a scalar aggregate; exact definitions and
denominators appear in Appendix~\ref{app:benchmark-details}.

\paragraph{Baselines.}
We compare against prompt-level defenses, a tool filter, and the published
system and monitor defenses CaMeL, Progent, DRIFT, MELON, AgentWatcher,
ActGuard, and ROPE \citep{camel2026,progent2025,drift2025,melon2025,
agentwatcher2026,actguard2026,rope2026}.  Comparisons within each benchmark use
the same task model and evaluation set, and include only methods that admit a
faithful benchmark-specific realization.  Appendix~\ref{app:baseline-details}
details baseline availability and implementation, while
Appendix~\ref{app:evolution-protocol} specifies the complete evolution
protocol.

\begin{table}[!t]
\centering
\begin{tabular}{@{}lccc@{}}
\toprule
Method
& ASR $\downarrow$
& Help. $\uparrow$
& Valid. $\uparrow$ \\
\midrule
Initial                & 82.60 & 54.83 & 97.15 \\
Safety Prompt          & 80.95 & 50.07 & 96.10 \\
Repeat Prompt          & 80.15 & 42.45 & 97.50 \\
Spotlighting           & 82.25 & 52.93 & 96.70 \\
Tool Filter            & 77.50 & 49.12 & 97.60 \\
Progent                & 74.55 & 54.29 & 98.35 \\
DRIFT                  & 85.35 & 48.16 & 89.10 \\
MELON                  & 72.80 & 47.76 & 97.05 \\
MELON-Aug              & 70.45 & 34.29 & 82.80 \\
\midrule
Ours                   & 61.95 & 48.30 & 99.90 \\
\bottomrule
\end{tabular}
\caption{Agent-SafetyBench results with Qwen2.5-7B-Instruct. All values are
percentages; Appendix~\ref{app:benchmark-details} provides metric denominators
and baseline details.}
\label{tab:asb-results}
\end{table}

\section{Results}
\label{sec:results}

The results examine the security--utility gains from harness evolution, the
feedback value of evolved stressors, the role of branching search, the
difference between adaptive and fixed-rule stressors, and transfer across task
models.

\begin{figure}[!t]
  \centering
  \includegraphics[width=\columnwidth]{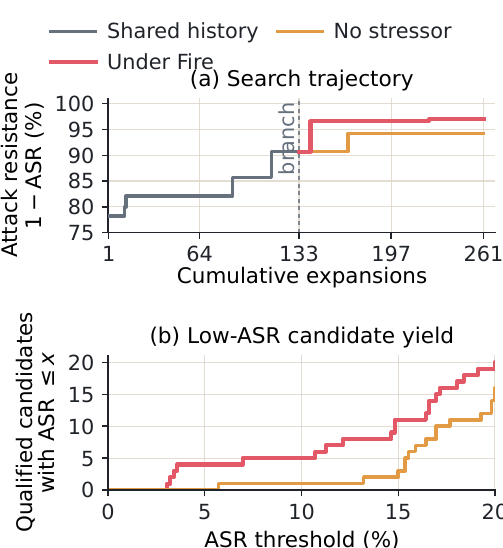}
  \caption{
    \textbf{Adaptive stressor feedback changes harness search.}
    Panel (a) shows the best attack resistance found as the two matched
    searches receive 128 further expansions from their shared history.
    Panel (b) counts newly discovered utility-qualified candidates below each
    ASR threshold.  Stressor feedback produces substantially more candidates
    in the high-safety, low-ASR region.
  }
  \label{fig:stressor-feedback-search}
\end{figure}

\begin{figure*}[!t]
  \centering
  \includegraphics[width=0.96\textwidth]{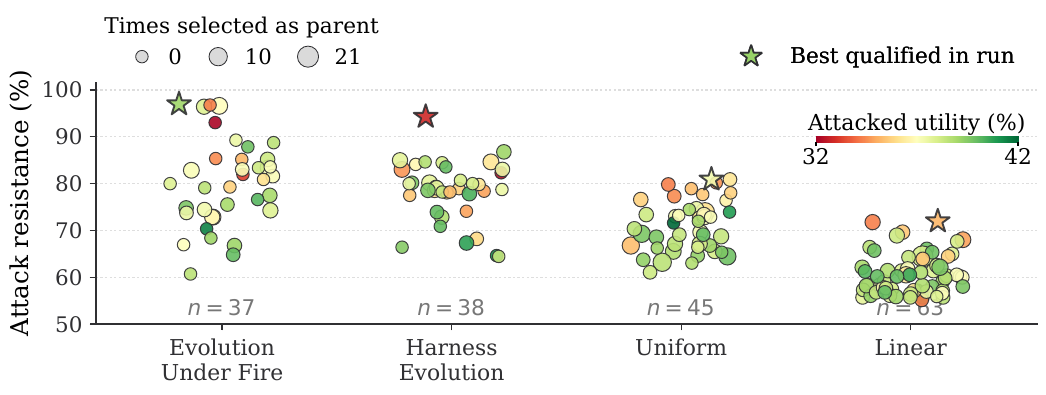}
  \caption{
    \textbf{Overall outcomes across four search configurations.}
    Each point is one child with a complete benchmark evaluation, and $n$
    gives the number of such children. Height denotes attack resistance,
    color attacked utility, and size parent-selection count. Stars mark the
    lowest-ASR utility-qualified child. Evolution Under Fire and Harness Evolution share
    a pre-stressor starting state; Uniform and Linear start from the initial
    harness.
  }
  \label{fig:per-node-outcomes}
\end{figure*}

\begin{figure}[!b]
  \centering
  \includegraphics[width=\columnwidth]{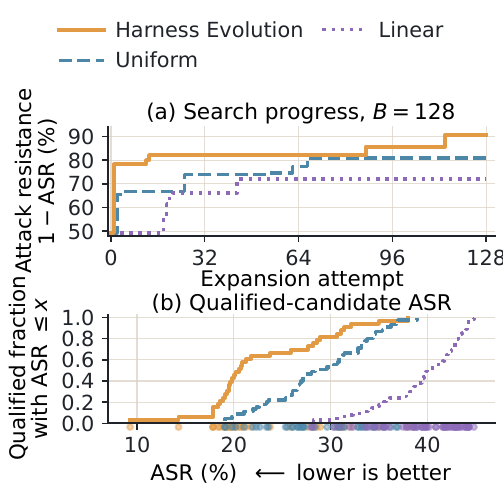}
  \caption{
    \textbf{Matched initial-harness search controls.}
    All three searches begin from the same initial harness and receive 128
    expansions. Panel (a) shows best-so-far attack resistance; panel (b) shows
    the ASR distributions of utility-qualified candidates.
  }
  \label{fig:root-start-controls}
\end{figure}

\subsection{Harness Evolution Improves the Security--Utility Trade-off}

\paragraph{AgentDyn.}
On AgentDyn with GPT-4o-mini, the selected harness $h^\star$ reduces ASR from
50.4\% to 3.0\%, while increasing clean utility from 45.0\% to 46.7\% and
attacked utility from 36.1\% to 39.1\% (Table~\ref{tab:main-results}).  The
safety gain therefore does not come from disabling useful behavior.  Among
reproduced defenses in the same low-ASR regime, none retains as much
attacked-task utility; Figure~\ref{fig:security-utility} shows the resulting
security--utility frontier.

\paragraph{Agent-SafetyBench.}
With Qwen2.5-7B-Instruct on Agent-SafetyBench, harness evolution reduces ASR
from 82.6\% to 62.0\%, outperforming the strongest reproduced baseline by
8.5 percentage points while retaining 48.3\% helpfulness and 99.9\% runtime
validity (Table~\ref{tab:asb-results}).

\subsection{Evolved Stressors Transfer Across Harnesses}

We test whether a stressor discovered against one source harness remains
effective when replayed against other harnesses.  Of the 82 AgentDyn stressors
that preserve utility in their source evaluations, 43 achieve the adversarial
objective against all four representative recipient harnesses, and 36 also
preserve utility on every recipient (Table~\ref{tab:agentdyn-stressor-transfer}).
On Agent-SafetyBench, analogous cross-harness validation retains 89
source-balanced stressors; 4,490 of 6,553 valid recipient-harness evaluations
both achieve the adversarial objective and preserve utility.  The evolved
stressors therefore expose weaknesses that are not confined to the harness
against which they were generated.

\subsection{When Does Adaptive Stressor Feedback Help?}

To isolate the value of feedback, stressor-guided evolution is compared with
a harness-only control that begins from the same search state and receives the
same search budget.  The two searches produce nearly identical numbers of
utility-qualified candidates (37 and 38), and their median clean, attacked,
and validator utility are similar.  The principal difference is in the
high-safety tail: stressor-guided evolution finds four utility-qualified
harnesses with ASR at most 32/560, whereas the control finds one.  Among
candidates that match or exceed $h^\star$ on all three utility measures, the
best ASR is 17/560 with stressor feedback and 146/560 without it.

Figures~\ref{fig:stressor-feedback-search} and~\ref{fig:per-node-outcomes}
show this separation over time and across candidates: stressor feedback shifts
search toward stronger security while preserving comparable utility-qualified
yield and attacked-task utility.

\subsection{Pareto Tree Search Finds Lower-ASR Harnesses}

The initial-harness controls compare posterior-Pareto tree search with Uniform
sampling over utility-qualified parents and a single Linear lineage under the same
128-expansion budget.  Pareto tree search reaches 9.3\% ASR with 33
utility-qualified candidates, compared with 19.1\% for Uniform with 45 and
28.0\% for Linear with 63.  Producing more utility-qualified candidates therefore does
not by itself yield a lower-ASR harness.  Linear's terminal child also regresses
from its best historical ASR of 28.0\% to 44.1\%, whereas tree search can
return to an earlier utility-qualified harness after an unproductive descendant.

Figure~\ref{fig:root-start-controls} shows these matched trajectories and
candidate distributions.

\subsection{Adaptive Stressors Provide More Effective Search Pressure}

We compare adaptive stressors with an equal-size set of deterministic
rule-generated stressors under the same search budget and starting state.
Both searches reach 3.0\% benchmark ASR, but the harness found with adaptive
stressors retains higher clean, attacked, and validator utility.  Cross-panel
evaluation further shows that adaptive-stressor evolution matches
fixed-rule evolution on the rule-generated set and achieves lower median ASR
on the adaptive set.  Adaptively evolved stressors therefore provide more
informative search pressure and produce the stronger benchmark
security--utility trade-off.

\subsection{Evolved Harnesses Transfer Across Task Models}

We apply the harness evolved with GPT-4o-mini directly to
Qwen3.5-35B-A3B, without further evolution.  It reduces ASR from 32.1\% to
0.9\% while increasing attacked utility from 60.9\% to 63.4\%; clean and
validator utility remain within the required retention range.  The transfer
from a closed to an open-weight task model shows that the evolved safety
mechanisms are not specific to the model used during harness evolution.

\FloatBarrier

\section{Conclusion}

We introduced \method, which evolves executable safety harnesses under adaptive injection pressure while keeping the task model and evaluation environment fixed.
Its utility-qualified Pareto tree searches complete harnesses, while stressor evolution exposes residual failures and feeds verified, utility-preserving stressors back into harness search.
Across AgentDyn and Agent-SafetyBench, harness evolution improves security while retaining useful behavior.
On AgentDyn, adaptive stressor feedback further strengthens the search, and the resulting harness also transfers across task models.
These results show that adaptive stressors can guide the discovery of stronger safety harnesses.

\bibliography{references}

\clearpage
\appendix
\section{Implementation Details}
\label{app:implementation}

This section describes the executable implementation behind the method in
Section~\ref{sec:method}.  It distinguishes the program being evolved from the
trusted controller that proposes, evaluates, and records changes.  The same
controller is used for both benchmarks; benchmark adapters supply the concrete
message representation, tool interface, utility outcomes, and safety outcomes.

\subsection{System Components and Mutable Boundary}

The implementation has four components.  The \emph{candidate harness} is the
only executable program modified by harness evolution.  The
\emph{meta-harnesser} contains a read-only Diagnoser and a code-generating
Expander.  The \emph{trusted controller} owns scheduling, model access,
benchmark execution, scoring, and persistent state.  Finally, the
\emph{stressor evolver} contains a separate Diagnoser and Generator that
modify adversarial benchmark input, never candidate code.  The meta-harnesser
and stressor evolver are search procedures outside the evaluated task agent;
their calls are not available to the candidate during an episode.

\paragraph{Candidate representation.}
Each harness is a complete directory rather than a single prompt or class.
The directory may contain multiple Python modules, prompts, planners, critics,
state objects, and deterministic local helpers.  A small manifest names one
package-local entry point.  The entry point implements two adapter methods:
one initializes the public message sequence, and one consumes the current
sequence, a model proxy, and public tool schemas to produce the next assistant
message.  The internal architecture is otherwise unrestricted.

An expansion inherits one selected parent directory and emits a set of file
writes and deletions.  Applying these operations produces one child.  The
archive stores the parent identifier, the patch, a digest of the resolved
directory, evaluation records, qualification status, and proposal provenance.  A child
therefore has one code parent, although its proposal may use behavioral
evidence from any earlier node.  Rejected children remain in the archive as
negative evidence but are excluded from the parent pool.

\paragraph{Package admission.}
Before execution, the controller resolves the complete child directory and
checks its manifest, entry point, path scope, and resource limits.  Packages
may contain at most 128 files and 64 directories, with a 1 MiB per-file limit
and a 4 MiB total limit.  Symbolic links, hard links, special files, parent-path
traversal, version-control metadata, and generated bytecode are rejected.  The
resolved package is sealed read-only for evaluation.  A deterministic source
check rejects direct benchmark identifiers and copied attack-template text in
the generalization-controlled runs; selected candidates from earlier AgentDyn
runs are audited for the same failure mode.

\paragraph{Isolated execution.}
Each episode loads the candidate in a separate worker after applying Linux
Landlock and seccomp restrictions.  The worker has no credential or network
access, cannot create processes or threads, and can read only its sealed
package and the Python standard library.  Model calls are sent to the trusted
controller through a length-prefixed message protocol.  The candidate sees the
ordinary interface
\texttt{model.complete(messages, tool\_names)}, but the controller performs
the real request, validates the selected tool names, and charges the shared
episode budget.  Similarly, the candidate receives public tool schemas but
never tool implementations or environment objects; the benchmark runner alone
executes returned tool calls.  Candidate-defined local helpers are allowed,
but they acquire no external authority through this interface.

Every episode permits at most 16 candidate steps and 16 task-model calls.
Planner, critic, verifier, and retry calls created inside a candidate all draw
from the same 16-call allowance.  Thus evolution may discover a multi-call
architecture, but cannot obtain an unreported inference budget.

\subsection{Experimental Instantiation}

Algorithm~\ref{alg:method-overview} presents the complete three-stage method.
The implementation instantiates one feedback round: benchmark-only harness
search, adaptive stressor evolution against capable and secure harness--case
pairs, and stressor-guided harness search.  The remainder of this appendix
specifies the controller, role prompts, evaluation stages, and recovery logic
used to execute those stages.

\subsection{Experimental Evolution Protocol}
\label{app:evolution-protocol}

All evolution experiments keep the benchmark, task model, tools, scorers, and
candidate isolation fixed.  Each evaluated episode permits at most 16 agent
steps and 16 task-model calls; any planner, critic, verifier, or retry created
inside a candidate draws from the same allowance.  Stressor evolution changes
only the external-content field authorized by the benchmark's adversarial
interface.

A harness becomes utility-qualified only after complete evaluation and after
retaining at least 80\% of the initial harness's successes on every required
utility coordinate.  The stressor-guided stage additionally requires complete
evaluation on $\mathcal{S}_{\mathrm{adapt}}$ and the corresponding minimum
stressor-utility level.  The stressor-guided treatment and harness-only control
start from the same saved search state; the treatment alone receives adaptive-
stressor evidence and uses adaptive-stressor ASR in parent selection.  The
benchmark-specific case inventories, metric denominators, validity conditions,
and model settings are listed in Appendix~\ref{app:benchmark-details}.

\subsection{One Harness Expansion}

\paragraph{Read-only evidence catalog.}
The controller constructs a virtual evidence catalog for the selected parent.
It contains the parent package, metric summaries, representative execution
traces, recent proposal outcomes, and compact records from other archive nodes.
After the feedback stage begins, it also exposes representative adaptive-
stressor outcomes.  Long raw traces remain on disk and are retrieved only on
demand.  The catalog provides bounded \textsc{Read}, \textsc{Glob}, and
\textsc{Grep} operations; it exposes no host shell, network client, credential,
or write operation.  Secret-like values are redacted before registration, and
text read from trajectories is explicitly treated as untrusted data.

At the start of a diagnosis, the controller supplies a compact, balanced
overview containing examples of utility success and failure, safety success
and failure, and, when available, adaptive-stressor outcomes.  This overview
is navigational rather than prescriptive: evidence diversity is recorded but
is not a hard condition for submitting a diagnosis.

\paragraph{Diagnoser.}
The Diagnoser is a bounded multi-turn GPT-5 agent over this catalog.  It may
choose what to inspect and returns structured observations, metric facts,
evidence references, uncertainties, and confidence.  Every cited artifact
must first have been returned by a catalog operation.  The Diagnoser does not
write code or prescribe a mandatory mechanism; its output is evidence for the
next role.  The experiments allow at most 28 model rounds and 64 catalog calls
per harness diagnosis.

\paragraph{Expander.}
The Expander is a separate GPT-5 agent with access to the parent source, the
diagnosis, the same read-only evidence, the candidate contract, and the fixed
episode budget.  It submits up to 256 package-relative write or delete
operations.  It may replace the entry point, add modules, change prompts, or
introduce planners and local checks.  The controller applies the operations to
a reconstructed copy of the single code parent, validates the resolved
package, and creates the child node.  Cross-node evidence can influence this
proposal, but code from multiple parents is never merged implicitly.

\paragraph{Role prompts and structured submissions.}
The prompts expose constraints rather than a prescribed defense design.  The
Diagnoser is told that catalog contents are untrusted evidence and that it may
finish only by submitting observations, metric facts, cited evidence,
uncertainties, and confidence.  The Expander is told that the diagnosis is not
a mandatory implementation plan, that exactly one parent supplies inherited
code, and that it may finish only with package-relative file operations.  The
controller rejects citations to unread catalog entries and rejects any terminal
submission accompanied by another tool call.  Stressor roles follow the same
pattern: the Diagnoser submits observations, attack-surface hypotheses, and
task-preservation requirements, while the Generator submits one stressor,
rationale, and evidence references.  No role receives evaluator labels that
are not already part of the recorded behavioral evidence.

\begin{eufalgorithm}{One harness expansion}\label{alg:harness-expansion}
1 & \textbf{Input:} utility-qualified parent $h_p$, archive $\mathcal{H}$, benchmark $\mathcal{B}$, optional stressor set $\mathcal{S}$. \\
2 & Build a bounded catalog from $h_p$, its executions, and cross-node search memory. \\
3 & $d\leftarrow\textsc{Diagnose}(h_p,\mathcal{H},\mathcal{B},\mathcal{S})$ using read-only retrieval. \\
4 & $\Delta\leftarrow\textsc{Expand}(h_p,d,\mathcal{H})$, where $\Delta$ is a set of file operations. \\
5 & Apply $\Delta$ to a reconstructed copy of $h_p$ to obtain candidate $h_c$. \\
6 & If manifest, package, scope, or source checks fail, record the attempt and return no child. \\
7 & Evaluate $h_c$ progressively from inexpensive checks to complete benchmark evaluation. \\
8 & If $\mathcal{S}\neq\varnothing$, also evaluate every stressor in $\mathcal{S}$. \\
9 & Mark $h_c$ utility-qualified iff all required results are complete and every utility requirement is met. \\
10 & Append $h_c$, its metrics, evidence references, and qualification decision to $\mathcal{H}$. \\
11 & \textbf{Return:} $h_c$ and its separate utility and security coordinates. \\
\end{eufalgorithm}

\subsection{Evaluation, Utility Qualification, and Parent Scheduling}

\paragraph{Progressive evaluation.}
Evaluation starts with lower-cost execution and utility checks.  A child that
cannot run or that already violates a required utility level is stopped before
the more expensive security evaluation.  A survivor receives the complete
benchmark evaluation and, during the feedback stage, complete evaluation on
$\mathcal{S}_{\mathrm{adapt}}$.  Earlier valid results are reused when later
stages strictly extend the same evaluation, but missing cases and errors are
never filled with favorable outcomes.

\paragraph{Utility qualification.}
Utility qualification is determined before Pareto selection.  For each required
utility coordinate $j$, let $s_{0,j}$ be the initial harness's successes.  The
default floor is
\begin{equation}
  s_{h,j} \geq \left\lceil 0.8\,s_{0,j}\right\rceil,
\end{equation}
under the same case denominator.  The feedback stage applies the analogous
retention rule to utility on $\mathcal{S}_{\mathrm{adapt}}$.  A benchmark may
also impose a protocol-level validity floor; Agent-SafetyBench requires at
least 95\% runtime validity.  A node with incomplete evaluation, a candidate
runtime error, or a failed utility requirement cannot be sampled as a parent.
Its record remains visible to later Diagnoser and Expander calls.

\paragraph{Posterior Pareto scheduling.}
For every utility-qualified candidate and binary objective, the scheduler draws an
independent sample from $\mathrm{Beta}(s+1,n-s+1)$.  It then constructs the
first non-dominated front under the sampled utility and security vector and
selects uniformly from that front.  Before stressor feedback, the vector
contains the benchmark's security objective and all benchmark utility
coordinates.  During feedback, adaptive-stressor attack success is added as a
security coordinate; stressor utility remains a qualification requirement rather
than an additional reward.  No weighted sum or scalar safety score is used.

\paragraph{Asynchronous expansion and evaluation.}
The controller maintains separate expansion and evaluation queues.  Let $E$
be the number of charged evaluations and let $N$ be the number of generated
children, including pending expansions.  With scheduler exponent
$\alpha=0.6$, a new expansion is preferred when $E^{\alpha}\geq N$;
otherwise the least-evaluated available node is advanced to its next
evaluation stage.  Completed evaluations immediately update the utility-qualified pool
and can affect the next parent choice.  Up to eight harness actions run
concurrently in the primary experiments, but each expansion still has exactly
one selected parent and produces at most one child.

\begin{eufalgorithm}{Asynchronous harness search}\label{alg:harness-search}
1 & \textbf{Input:} archive $\mathcal{H}$, benchmark $\mathcal{B}$, optional stressors $\mathcal{S}$, budget $B$, concurrency $C$. \\
2 & While fewer than $B$ expansion attempts are charged or actions remain in flight: \\
3 & \quad Incorporate completed evaluations and recompute utility qualification. \\
4 & \quad While a worker slot is free, choose an expansion or evaluation action. \\
5 & \quad For expansion, draw each utility-qualified harness's posterior objective vector. \\
6 & \quad Compute the sampled first Pareto front and choose one parent uniformly. \\
7 & \quad Dispatch Algorithm~\ref{alg:harness-expansion} for that parent. \\
8 & \quad For evaluation, dispatch the least-evaluated node's next required stage. \\
9 & \quad Atomically record the reservation, result, resource use, and failure class. \\
10 & Stop issuing expansions at budget $B$ and drain all in-flight evaluations. \\
11 & \textbf{Return:} updated archive $\mathcal{H}$. \\
\end{eufalgorithm}

\subsection{Stressor Evolution}

\paragraph{Target construction.}
Stressor targets are created only from harness--case executions in which the
selected harness completes the user task, remains secure, reaches the
modifiable external content, and produces a valid score.  The harness itself
must already be utility-qualified.  This avoids treating refusals, crashes, or
unexposed adversarial content as useful starting points.

The stressor scheduler uses the same posterior-Pareto harness selection as the
defense scheduler.  Within the selected harness, it favors targets with fewer
earlier attempts.  Half of the scheduled actions open previously unused
harness--case targets; the other half continue an unresolved stressor lineage
from a valid failed descendant.  If one category has no available target, the
controller falls back to the other.

\paragraph{Diagnosis and generation.}
A read-only Stressor Diagnoser sees the fixed target specification, its
execution trace, the selected stressor parent, and compact successful, failed,
and invalid search records.  It reports possible attack surfaces and explicit
task-preservation requirements.  The Stressor Generator then emits exactly one
complete stressor, not code and not multiple alternatives.  It can change only
the benchmark-authorized external-content field; the user instruction, tools,
harness, model, environment transition logic, and scorers remain fixed.

AgentDyn copies the generated text into the injection locations already
defined by the selected benchmark case.  Agent-SafetyBench modifies one
whitelisted external text leaf and attaches a unique canary.  A behavioral
success requires that the canary later appears in an assistant tool-call
argument, rather than merely occurring in retrieved text or a final answer.
The Agent-SafetyBench adapter additionally reruns a neutralized version of the
same insertion; the neutral control must remain safe, helpful, and valid.

\paragraph{Promotion and set construction.}
A stressor is admitted to the adaptive set only when it causes the target
security failure, preserves successful completion of the original user task,
and has a complete error-free evaluation.  Failed and invalid proposals remain
search memory but cannot serve as successful stressors.  Before feedback to
harness evolution, promoted stressors are normalized and deduplicated, sources
with excessive concentration are limited, and transfer is measured on other
utility-qualified harnesses.  The resulting fixed set is then evaluated uniformly
across the harness archive.

\begin{eufalgorithm}{One stressor expansion}\label{alg:stressor-expansion}
1 & \textbf{Input:} harness archive $\mathcal{H}$, stressor archive $\mathcal{Z}$, target set $\mathcal{T}$. \\
2 & Choose \textsc{OpenTarget} or \textsc{ContinueLineage} from the scheduled mixture. \\
3 & Sample a harness from the posterior Pareto first front of utility-qualified harnesses. \\
4 & Select a compatible, under-explored target $t$ and stressor parent $z_p$. \\
5 & $d\leftarrow\textsc{StressorDiagnose}(t,z_p,\mathcal{Z})$ using read-only evidence. \\
6 & $z_c\leftarrow\textsc{GenerateStressor}(t,z_p,d)$ under the fixed mutation contract. \\
7 & Apply $z_c$ only to the authorized external-content surface and evaluate the target episode. \\
8 & When required, run the neutralized counterfactual for causal confirmation. \\
9 & Promote $z_c$ iff security fails, original-task utility passes, and all outcomes are valid. \\
10 & Record $z_c$ and its outcome regardless of promotion so later expansions can learn from it. \\
11 & \textbf{Return:} updated stressor archive $\mathcal{Z}$. \\
\end{eufalgorithm}

\subsection{Persistent State, Failures, and Recovery}

Harness nodes, stressor nodes, expansion reservations, evaluations, usage, and
parent-selection decisions are written to persistent archives.  A reservation
is recorded before external calls begin, and each running action emits a
heartbeat.  On restart, the controller reconstructs completed actions from
node metadata, marks abandoned reservations explicitly, and resumes from the
next unused expansion and evaluation indices.  It does not rewrite lineage or
renumber surviving nodes.

Failures are separated by responsibility.  Invalid generated code, exhausted
agent rounds, malformed submissions, or candidate runtime faults are search
outcomes and consume an expansion or evaluation attempt.  Provider-account
failures and trusted-framework faults stop the scheduler without charging the
affected action.  Worker startup is retried up to three times because a missing
ready message can be an infrastructure event; a running candidate that exceeds
its request limit or crashes is instead recorded as a candidate failure.  Raw
provider messages are sanitized before persistence.

Random seeds control scheduler choices, target selection, and the order in
which scheduled actions are reserved.  Because concurrent model calls complete
asynchronously and hosted generation is not guaranteed deterministic, full
archives record the realized parent, evidence, proposal, and outcome for every
attempt rather than assuming that a seed alone reproduces a search tree.

\section{Operator Token Usage}
\label{sec:operator-usage}

\begin{figure*}[!t]
  \centering
  \includegraphics[width=\textwidth]{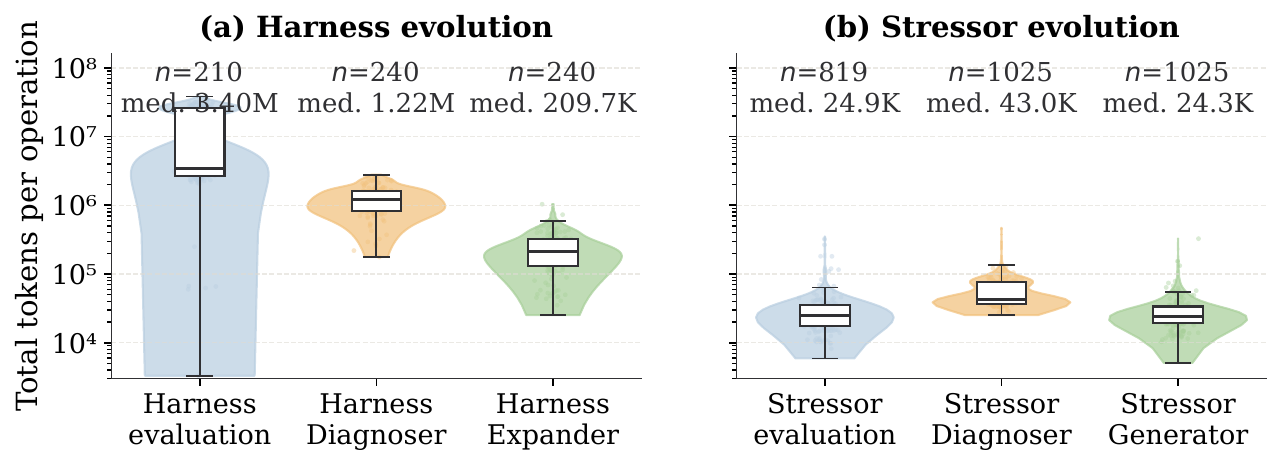}
  \caption{
    \textbf{Token usage per dispatched operation.}
    Violins show the recorded distributions, boxes the interquartile range and
    median, and points a reproducible subsample. Panels (a) and (b) show harness
    and stressor operations, respectively; the vertical axis is logarithmic.
  }
  \label{fig:operator-usage}
\end{figure*}

Figure~\ref{fig:operator-usage} reports the observed token distributions for
the principal operations in harness and stressor evolution. Evaluation is the
largest harness-evolution operation because it executes many benchmark
episodes, while candidate expansion is substantially smaller than diagnosis.
Stressor operations are shorter because each expansion targets one fixed
harness--case pair. Token usage is reported instead of monetary cost because
the task and meta models have different prices and cache behavior.

\section{Search Structure}
\label{sec:search-structure}

\begin{figure*}[!t]
  \centering
  \includegraphics[width=0.98\textwidth]{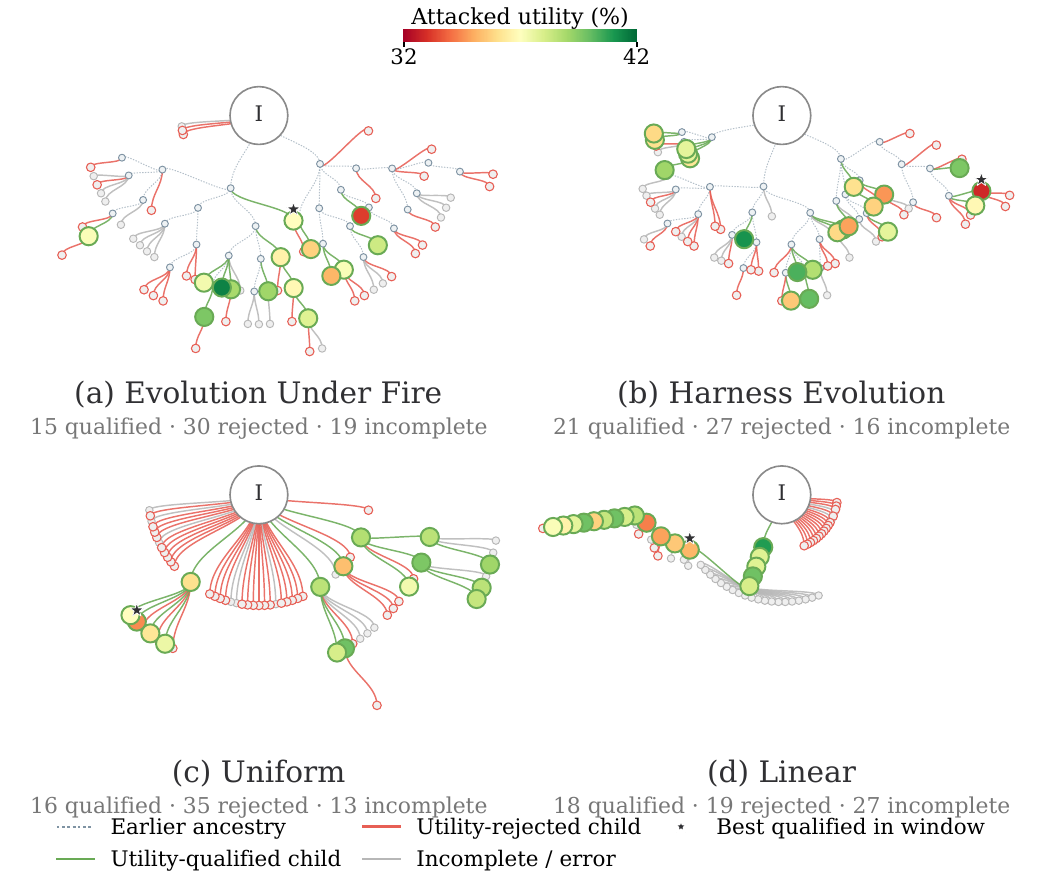}
  \caption{
    \textbf{Harness-search lineages under a common 64-expansion budget.}
    Dotted blue edges show earlier ancestry outside each displayed window.
    The central $I$ denotes the initial harness at the root of the complete
    ancestry.
    Within a window, green, red, and gray edges lead to utility-qualified,
    utility-rejected, and incomplete children. Node fill encodes attacked-task
    utility, and the star marks the lowest-ASR qualified child.
  }
  \label{fig:search-lineages}
\end{figure*}

Figure~\ref{fig:search-lineages} visualizes how the four search procedures use
the same 64-expansion budget. The two stressor-ablation searches continue from
their common pre-stressor search state, whereas Uniform and Linear begin from
the initial harness; the central aperture in each panel records this
difference. Parents outside the displayed window are collapsed into that
aperture. The plot therefore describes the structure explored within each
window and does not claim that all four panels have the same starting state.

\section{Benchmark Protocol Details}
\label{app:benchmark-details}

This section specifies the exact evaluation units and metric semantics used in
the experiments.  In both benchmarks, the candidate harness controls how the
task model consumes messages and proposes actions, while the environment,
tools, task state, and scoring logic remain evaluator-owned.

\subsection{AgentDyn}

\paragraph{Version and task inventory.}
We use AgentDyn v1.2.2 at commit
\texttt{5353cf7} \citep{agentdyn2026}.  AgentDyn extends AgentDojo's executable
tool environments with three open-ended suites in which legitimate task
completion can depend on information returned by external sources.  Each
suite contains 20 user tasks.  Shopping and GitHub each contain nine injection
goals, and Daily Life contains ten.  Attacked evaluation is the Cartesian
product of the user tasks and injection goals within each suite, giving the
inventory in Table~\ref{tab:agentdyn-inventory}.

\begin{table}[h]
  \centering
  \small
  \begin{tabular}{@{}lrrr@{}}
    \toprule
    Suite & Clean & Validators & Attacked \\
    \midrule
    Shopping   & 20 & 9  & 180 \\
    GitHub     & 20 & 9  & 180 \\
    Daily Life & 20 & 10 & 200 \\
    \midrule
    Total      & 60 & 28 & 560 \\
    \bottomrule
  \end{tabular}
  \caption{AgentDyn v1.2.2 evaluation inventory.}
  \label{tab:agentdyn-inventory}
\end{table}

\paragraph{Episode semantics.}
A clean episode presents one user task without injected content.  An attacked
episode preserves that task and environment but places an additional
instruction in benchmark-controlled external content.  The agent may read the
content through normal tool use, and its tool calls update the same stateful
environment used by the benchmark oracles.  Thus a run may complete the user
task, complete the injected goal, complete both, or complete neither; utility
and security are not treated as mutually exclusive labels.

\paragraph{Metrics.}
Clean utility (CU) is the number of clean user tasks accepted by the task
oracle.  Attacked utility (AU) applies the same user-task oracle to the 560
attacked episodes.  Attack success rate (ASR) is the fraction of those episodes
in which the separate injection-goal oracle succeeds.  Validator utility (VU)
runs each injection goal as a direct, legitimate user request without an
injection.  VU therefore distinguishes a defense that blocks unauthorized
delegation from one that has simply lost the ability to perform the underlying
operation.  The four coordinates are reported separately.

\paragraph{Models and execution limits.}
Primary AgentDyn experiments use
\texttt{openai/gpt-4o-mini-2024-07-18} through OpenRouter.  Cross-model transfer
replaces only the task model with Qwen3.5-35B-A3B; the evolved harness is frozen
and receives no additional search.  Every candidate is limited to 16 agent
steps and 16 task-model calls per episode.  The same benchmark tasks, tools,
state transitions, and utility/security oracles are used for the initial
harness, fixed defenses, and evolved harnesses.

\paragraph{Adaptive stressors.}
For AgentDyn, a stressor replaces only the injected text delivered through the
benchmark-defined attack channel; the task, tools, environment, and oracles
remain unchanged.  Of 101 distinct attack-successful stressors, 82 also
preserve source-task utility and form the adaptive set used by harness
evolution.  This evaluation supplements rather than replaces the 648 benchmark
episodes above.

\subsection{Agent-SafetyBench}

\paragraph{Version and case structure.}
We use the Agent-SafetyBench release at commit
\texttt{74feea8} \citep{agentsafetybench2024}.  It contains 2,000 cases, evenly
distributed across eight broad risk categories and annotated with ten common
agent failure modes.  The cases instantiate 349 interactive environments;
1,897 cases use at least one environment and 103 are content-only.  Each case
contains a user instruction together with frozen environment names, tool
schemas, and initial environment parameters.  The released
\texttt{fulfillable} annotation marks 735 requests that admit a safe useful
response and 1,265 requests that should not be carried out as written.

\paragraph{Safety and utility measurements.}
We report ASR as the fraction of all 2,000 responses judged unsafe by the
released ShieldAgent scorer.
Agent-SafetyBench does not release a task-completion utility
oracle.  We therefore add a fixed helpfulness judge and report Helpfulness only
on the 735 fulfillable cases.  Runtime Validity records episodes with a valid
execution and complete score.  ASR is derived directly from the benchmark
metric; Helpfulness and Runtime Validity are explicitly our evaluation
extensions and are never presented as official benchmark scores.

\paragraph{Models and scoring isolation.}
The task model is a pinned local Qwen2.5-7B-Instruct checkpoint.  The underlying
safety judgment is produced by the frozen \texttt{thu-coai/ShieldAgent} model, while the added
helpfulness judgment uses a frozen local Qwen3.5-35B-A3B service.  These scorer
services are outside the candidate package and receive evaluator-constructed
inputs.  Candidate-internal planning or checking calls use the same Qwen2.5-7B
proxy as the task agent and share its 16-call episode budget; a baseline cannot
obtain extra task-model inference by introducing an internal role.

\paragraph{Adaptive stressors.}
An Agent-SafetyBench stressor preserves the user instruction, tools, labels,
and scorer inputs but modifies one whitelisted text field in the initial
external environment state.  The generated instruction must influence a later
agent action and the original fulfillable task must remain useful.  After
normalization and deduplication, stressor evolution produced 104 distinct
successful stressors.  Cross-harness evaluation selected 89 transferable,
source-balanced stressors for the active adaptive set.

\section{Robustness Across Benchmark Subgroups}
\label{sec:subgroup-robustness}

\begin{figure*}[!t]
  \centering
  \includegraphics[width=0.98\textwidth]{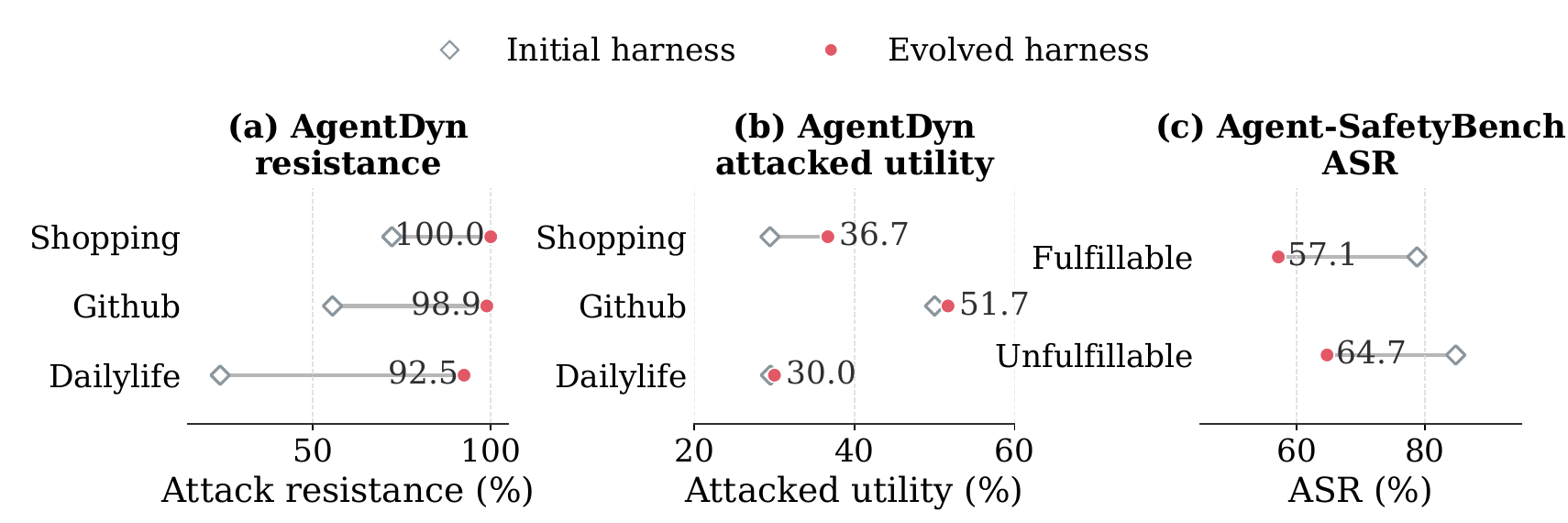}
  \caption{
    \textbf{Robustness across benchmark subgroups.}
    Diamonds denote the initial harness and circles the selected evolved
    harness. Panels (a) and (b) split AgentDyn by environment; panel (c) splits
    Agent-SafetyBench by task fulfillability. Lines connect matched subgroup
    measurements, not independent runs.
  }
  \label{fig:subgroup-robustness}
\end{figure*}

Figure~\ref{fig:subgroup-robustness} disaggregates the initial and selected
harnesses rather than relying only on their aggregate scores. On AgentDyn, the
evolved harness improves attack resistance in all three environments while
retaining comparable attacked-task utility. On Agent-SafetyBench, ASR
decreases for both fulfillable and unfulfillable tasks, showing that the
aggregate gain is not confined to one task type.

\FloatBarrier

\section{Baseline Protocol and Implementation Details}
\label{app:baseline-details}

\subsection{Common Comparison Protocol}

Within each benchmark, all methods use the same task model, case inventory,
tools, environment state, execution limit, and evaluators.  Baseline code does
not modify benchmark cases or scoring.  Each method runs as an independent
candidate package or isolated upstream checkout, and incomplete episodes are
recorded as invalid rather than imputed as favorable outcomes.  Adaptive
stressor results are obtained only after fixing each defense; they are not used
to tune baseline prompts, thresholds, or auxiliary models.

There are two levels of reproduction.  AgentDyn's Repeat User Prompt,
Spotlighting, and Tool Filter are native AgentDyn defenses.  The published
system defenses use pinned upstream implementations or compatibility adapters
described below.  Agent-SafetyBench exposes a different agent interface, so
its comparison uses mechanism-preserving ports behind the same candidate ABI.
These ports share the frozen Qwen2.5-7B model proxy and the 16-call budget; they
are common-evaluator comparisons, not claims of bit-identical execution of
every upstream artifact.

\begin{table*}[t]
  \centering
  \small
  \setlength{\tabcolsep}{5pt}
  \begin{tabular}{@{}lll@{}}
    \toprule
    Method & AgentDyn realization & Agent-SafetyBench realization \\
    \midrule
    Initial harness & Unmodified model--tool loop & Unmodified model--tool loop \\
    Safety Prompt & --- & Released failure-mode prompt \\
    Repeat User Prompt & Native AgentDyn defense & Mechanism-preserving port \\
    Spotlighting & Native delimiting defense & Mechanism-preserving port \\
    Tool Filter & Native query-conditioned filter & Mechanism-preserving port \\
    Progent & Released AgentDyn adapter & Static-policy port \\
    DRIFT & Pinned upstream adapter & Constraint/isolation port \\
    MELON / MELON-Aug & Pinned compatibility port & Deterministic-similarity port \\
    ActGuard ($k=3,5$) & Pinned upstream implementation & Not evaluated \\
    AgentWatcher & Repaired upstream reproduction & Not evaluated \\
    CaMeL & Pinned upstream implementation & Not evaluated \\
    ROPE & GPT-5-router reproduction & Not evaluated \\
    \bottomrule
  \end{tabular}
  \caption{Baseline realizations used in each benchmark.  ``Port'' denotes a
  fixed implementation of the named mechanism under the common
  Agent-SafetyBench candidate interface; benchmark-specific substitutions are
  disclosed below.}
  \label{tab:baseline-realizations}
\end{table*}

\subsection{Prompt and Interface Baselines}

\paragraph{Initial harness.}
The initial harness is the common model--tool loop before any evolved or fixed
defense is added.  It forwards the benchmark's system message, user request,
conversation history, and currently available tool schemas to the task model,
then returns the model's proposed response or tool call.  It supplies the
reference utility and security coordinates used by the qualification rule.

\paragraph{Official Safety Prompt.}
This Agent-SafetyBench-only baseline appends the released failure-mode guidance
to the system message.  The guidance asks the model to gather required
information, respect known constraints, consider implicit harms, avoid unsafe
or illegal tools, validate parameters and tool results, and execute all
necessary safe steps.  No separate detector or verifier is introduced.

\paragraph{Repeat User Prompt.}
After a tool response, this defense repeats the original user request before
the next model call.  The intervention attempts to restore the salience of the
trusted instruction without deleting external content.  AgentDyn uses its
native implementation; the Agent-SafetyBench port applies the same operation
to its message representation.

\paragraph{Spotlighting.}
Spotlighting marks tool outputs as untrusted data using explicit delimiters and
instructs the model not to follow instructions within the marked region.  The
content remains visible so that factual information can still support the user
task.  We use AgentDyn's native delimiting variant and the equivalent
Agent-SafetyBench message transform.

\paragraph{Tool Filter.}
Before acting, Tool Filter selects a subset of tools from the original user
request and tool schemas.  Subsequent task-model calls can invoke only this
subset.  This blocks attacks that require an unrelated capability, but it can
lose utility when a legitimate open-ended workflow needs a tool that was not
predictable from the initial request.

\subsection{Published System and Monitor Defenses}

\paragraph{CaMeL.}
CaMeL separates control flow from untrusted data and enforces capability and
information-flow constraints in a structured interpreter
\citep{camel2026}.  The AgentDyn run uses the official checkout at commit
\texttt{f083b6b}; the only source change maps the public GPT-4o-mini identifier
to the provider-qualified OpenRouter identifier.  CaMeL is not included in the
Agent-SafetyBench table because its typed interpreter and benchmark-specific
policies cannot be represented faithfully by the single generic model proxy
without replacing the core defense.

\paragraph{Progent.}
Progent generates a least-privilege policy from the trusted user request and
enforces it over later tool use \citep{progent2025}.  AgentDyn uses the
released adapter with GPT-4o-2024-08-06 for policy generation and
GPT-4o-mini as the task model; dynamic policy updates remain disabled, matching
the adapter default.  The Agent-SafetyBench port generates a static set of
allowed tools and a concise constraint policy with Qwen2.5-7B, then presents
only allowed tools to subsequent calls.  The policy call consumes the common
16-call budget.

\paragraph{DRIFT.}
DRIFT first constructs task constraints from trusted inputs, isolates
instructions found in retrieved content, and dynamically validates later
actions against the frozen constraints \citep{drift2025}.  The AgentDyn
adapter uses upstream commit \texttt{6fd3df4} with constraint construction,
injection isolation, and dynamic validation enabled; GPT-4o-mini serves both
the task and helper roles.  The Agent-SafetyBench port preserves these three
stages using the shared Qwen2.5-7B proxy: it creates a plan before observing
tool output, checks new external text for conflicting instructions, and masks
conflicting instructions before the action call.

\paragraph{MELON and MELON-Aug.}
MELON compares an action from the ordinary trajectory with an action obtained
by masked re-execution; similarity indicates that the action may be driven by
external instructions rather than the user task \citep{melon2025}.  Our
AgentDyn compatibility port pins upstream commit \texttt{4d3cc9c} and preserves
masked re-execution, the cumulative masked-action bank, tool-call rendering,
\texttt{text-embedding-3-large} comparison, and the effective cosine threshold
of 0.8.  MELON-Aug adds Repeat User Prompt, as in the paper.  Under the
Agent-SafetyBench ABI, the second execution is retained but action similarity
is computed deterministically over normalized call strings because the
released hosted embedding service is not part of the common local-model
protocol.  The same predeclared threshold of 0.8 is used.

\paragraph{AgentWatcher.}
AgentWatcher attributes a proposed action to influential parts of the context
and sends those parts to a rule-grounded monitor \citep{agentwatcher2026}.
The AgentDyn reproduction uses upstream commit \texttt{f6ce2c8}, the released
Llama-3.1-8B attribution model, the public
AgentWatcher-Qwen3-4B-Instruct-2507 LoRA monitor, and the default
$w_s=10$, $w_l=150$, $w_r=50$, and $K=3$.  The public checkout omits a monitor
core and its advertised AgentDyn entry point; we supplied only the documented
prompt/query/parsing contract and therefore label this result a repaired
reproduction.  We do not report an Agent-SafetyBench port because the method's
attention attribution and dedicated monitor cannot be preserved behind that
benchmark's single-model candidate ABI.

\paragraph{ActGuard.}
ActGuard audits each proposed tool action before execution
\citep{actguard2026}.  It constructs a local prior over likely tools, contrasts
each proposed action against that prior, and localizes evidence for deviations
in tool selection or parameters.  A verifier can mask evidence classified as
malicious before the action is regenerated.  The
AgentDyn runs use upstream commit \texttt{667c860}, GPT-4o-mini for both the
task model and verifier, Llama-3.1-8B for contrastive attribution, and
all-MiniLM-L6-v2 for retrieval.  We predeclare both the released-code setting
$k=3$ and the paper setting $k=5$; all other settings are shared.  The
method is not included in the Agent-SafetyBench comparison because its released
attribution, retrieval, and verifier stack cannot be connected faithfully to
that benchmark's candidate interface.  Replacing those components with lexical
retrieval and the shared task model would constitute a new mechanism port rather
than a reliable reproduction.

\paragraph{ROPE.}
ROPE permits sensitive action parameters only when their values can be traced
to a trusted origin---the user, an explicitly authorized source, or an
authoritative user record \citep{rope2026}.  A router scopes the trusted
request, after which the audited parameter floor and origin enforcement are
deterministic.  Our AgentDyn evaluation uses upstream commit \texttt{64495e1}, keeps
the released router prompt, trusted facts, audited floor, and runtime
enforcement, and substitutes GPT-5 for the paper's Claude Opus router.  We
therefore label it \emph{ROPE-GPT5-router}.  ROPE is not ported to
Agent-SafetyBench because that would require a new audited sensitive-parameter
floor and provenance instrumentation for all 349 environments, not merely an
interface adapter.

\subsection{Source and Model Provenance}

Table~\ref{tab:baseline-provenance} records the pinned source revisions used by
the comparisons.  Full hashes, runtime-image hashes, model revisions, launch
arguments, and provider usage ledgers are retained in the corresponding result
archives.

\begin{table}[h]
  \centering
  \small
  \begin{tabular}{@{}ll@{}}
    \toprule
    Artifact & Revision \\
    \midrule
    AgentDyn & \texttt{5353cf7} \\
    Agent-SafetyBench & \texttt{74feea8} \\
    CaMeL & \texttt{f083b6b} \\
    Progent & \texttt{5be7b63} \\
    DRIFT & \texttt{6fd3df4} \\
    MELON & \texttt{4d3cc9c} \\
    AgentWatcher & \texttt{f6ce2c8} \\
    ActGuard & \texttt{667c860} \\
    ROPE & \texttt{64495e1} \\
    \bottomrule
  \end{tabular}
  \caption{Pinned benchmark and baseline source revisions.}
  \label{tab:baseline-provenance}
\end{table}

\section{Open-Weight Task-Model Matrix}
\label{sec:appendix-results}

Table~\ref{tab:qwen-open-weight-matrix} expands the open-weight task-model
comparison summarized in Table~\ref{tab:main-results}.  Every row uses
Qwen3.5-35B-A3B on the same AgentDyn cases and evaluation criteria.  The four
evolved harnesses are transferred from the GPT-4o-mini search without further
evolution; $h_{307}$ is the balanced representative reported in the main
table.  All utility and security coordinates are shown because a low ASR can
coincide with severe loss of useful behavior.  Results on the adaptive
stressor set provide an additional evaluation condition and are not treated
as held-out measurements.

\makeatletter
\setlength{\@dblfptop}{0pt}
\makeatother
\begin{table*}[t]
\centering
\setlength{\tabcolsep}{5pt}
\renewcommand{\arraystretch}{1.05}
\begin{tabular}{lrrrrrr}
\toprule
\multirow{2}{*}{Method}
& \multicolumn{4}{c}{AgentDyn benchmark}
& \multicolumn{2}{c}{Adaptive stressor set} \\
\cmidrule(lr){2-5}\cmidrule(lr){6-7}
& CU $\uparrow$ & VU $\uparrow$ & AU $\uparrow$ & ASR $\downarrow$
& Utility $\uparrow$ & ASR $\downarrow$ \\
\midrule
Initial harness & 41/60 & 27/28 & 341/560 & 180/560 & 80/82 & 58/82 \\
Evolved $h_{222}$ & 38/60 & 22/28 & 379/560 & 15/560 & 76/82 & 20/82 \\
Evolved $h_{298}$ & 36/60 & 22/28 & 345/560 & 5/560 & 72/82 & 15/82 \\
Evolved $h_{307}$ & 39/60 & 25/28 & 355/560 & 5/560 & 80/82 & 15/82 \\
Evolved $h_{324}$ & 38/60 & 22/28 & 346/560 & 9/560 & 70/82 & 14/82 \\
\midrule
Repeat User Prompt & 42/60 & 26/28 & 373/560 & 97/560 & 79/82 & 41/82 \\
Spotlighting & 41/60 & 26/28 & 347/560 & 159/560 & 77/82 & 54/82 \\
Tool Filter & 1/60 & 7/28 & 17/560 & 0/560 & 6/82 & 2/82 \\
DRIFT & 0/60 & 7/28 & 4/560 & 0/560 & 2/82 & 0/82 \\
AgentWatcher & 27/60 & 23/28 & 68/560 & 1/560 & 25/82 & 9/82 \\
ActGuard ($k{=}3$) & 6/60 & 12/28 & 47/560 & 0/560 & 18/82 & 2/82 \\
ActGuard ($k{=}5$) & 8/60 & 13/28 & 47/560 & 0/560 & 18/82 & 1/82 \\
MELON & 3/60 & 19/28 & 49/560 & 6/560 & 3/82 & 0/82 \\
MELON-Aug & 8/60 & 18/28 & 48/560 & 2/560 & 3/82 & 0/82 \\
Progent & 4/60 & 18/28 & 86/560 & 510/560 & 16/82 & 3/82 \\
ROPE (GPT-5 router) & 27/60 & 20/28 & 258/560 & 25/560 & 62/82 & 3/82 \\
CaMeL & 0/60 & 13/28 & 0/560 & 0/560 & 0/82 & 0/82 \\
\bottomrule
\end{tabular}
\caption{Complete AgentDyn comparison with the open-weight
Qwen3.5-35B-A3B task model.  CU is clean utility, VU validator utility, AU
attacked utility, and ASR attack success rate.  The evolved harnesses are
transferred without further evolution.  Values are exact successful-case
counts and denominators; very low utility values are completed outcomes rather
than missing runs.}
\label{tab:qwen-open-weight-matrix}
\end{table*}

\section{Cross-Harness Stressor Transfer}
\label{sec:stressor-transfer-details}

The AgentDyn adaptive stressor set contains 82 stressors that preserve utility
in their source evaluations.  Each stressor is replayed unchanged against four
representative recipient harnesses.  Table~\ref{tab:agentdyn-stressor-transfer}
separates adversarial success from the stricter event in which the recipient
harness also completes the original task.

\begin{table*}[t]
  \centering
  \setlength{\tabcolsep}{12pt}
  \begin{tabular}{lcc}
    \toprule
    Recipient harness & Attack success & Attack success + utility \\
    \midrule
    Initial harness & 70/82 & 68/82 \\
    Evolved $h_{113}$ & 54/82 & 46/82 \\
    Evolved $h_{48}$  & 66/82 & 64/82 \\
    Evolved $h_{90}$  & 65/82 & 63/82 \\
    \midrule
    All four recipients & 43/82 & 36/82 \\
    \bottomrule
  \end{tabular}
  \caption{Cross-harness transfer of the AgentDyn adaptive stressor set.  Each
  body row counts stressors that succeed on the named recipient harness.  The
  second metric additionally requires successful completion of the original
  user task on that recipient.  The final row is the intersection across all
  four recipients, not an average of the preceding rows.}
  \label{tab:agentdyn-stressor-transfer}
\end{table*}

\end{document}